# PERSONAL INFORMATION PRIVACY SETTINGS OF ONLINE SOCIAL NETWORKS AND THEIR SUITABILITY FOR MOBILE INTERNET DEVICES

Nahier Aldhafferi, Charles Watson and A.S.M Sajeev

School of Science and Technology, University of New England, Australia

naldaffe@myune.edu.au
cwatson7@une.edu.au
sajeev@une.edu.au

## ABSTRACT

*Protecting personal information privacy has become a controversial issue among online social network providers and users. Most social network providers have developed several techniques to decrease threats and risks to the users' privacy. These risks include the misuse of personal information which may lead to illegal acts such as identity theft. This study aims to measure the awareness of users on protecting their personal information privacy, as well as the suitability of the privacy systems which they use to modify privacy settings. Survey results show high percentage of the use of smart phones for web services but the current privacy settings for online social networks need to be improved to support different type of mobile phones screens. Because most users use their mobilephones for Internet services, privacy settings that are compatible with mobile phones need to be developed. The method of selecting privacy settings should also be simplified to provide users with a clear picture of the data that will be shared with others. Results of this study can be used to develop a new privacy system which will help users control their personal information easily from different devices, including mobile Internet devices and computers.*

## 1. INTRODUCTION

Information and communication technology (ICT) plays a significant role in today's networked society. It has affected the online interaction between users, who are aware of security applications and their implications on personal privacy. There is a need to develop more security mechanisms for different communication technologies, particularly online social networks. Privacy is essential to the design of security mechanisms. Most social networks providers have offered privacy settings to allow or deny others access to personal information details. For example, MySpace website allows privacy settings for people under 18 years to make their profile content available only for their friends and others who are less than 18 years old. In addition to the expansion of using mobile Internet devices, some social networks websites have provided users with mobile versions to support different types of mobile phone screens.

Therefore, this article will study the relationship between personal information privacy settings of online social networks and their suitability for mobile Internet devices.

## 2. LITERATURE REVIEW

### 2.1. Privacy

The increased use of information and communication technologies has had a significant impact on the interactions between users. This is particularly true for people who use mobile devices to communicate with one another or to access the Internet. Mobile web users have difficulty knowing where and how their information is stored and who is authorised to use it. Therefore, protecting mobile web users' data and increasing their confidence in data privacy has become a real challenge.

Cavoukian (2009) coined the term "privacy by design" which refers to the need to address privacy concerns from the outset. The author defined it as a philosophy to enhance design by embedding privacy concerns as requirements into areas of design such as technology design, business practices and physical design.

The "privacy by design" issue has become the main basis of online application designs. Different social network providers such as Facebook, Google and Twitter are competing to provide a level of privacy in their applications that would inspire confidence in their users. Using the concept "privacy by design" as a standard for designing applications will give users more authority to decide what kind of information they want to share with whom.

#### 2.1.1 Definitions of privacy

Different definitions of privacy exist in the context of ICT. Bünnig and Cap (2009) described privacy as protecting personal information from being misused by malicious entities and allowing certain authorised entities to access that personal information by making it visible to them. Ni et al. (2010) defined privacy as set of privacy policies that force the system to protect private information. Taheri et al. (2010) claimed that privacy is especially important in a wide range of applications that seek to protect the user's location information and hide some details from others. Since the concept of privacy is diverse, no single definition of privacy encompasses all aspects of the term. So, based on the definition of privacy by Bünnig et al. (2009), this study is concerned primarily with information privacy.

### 2.2. Online personal information privacy

Protecting the privacy of personal information is one of the biggest challenges facing website developers, especially social network providers. Several researchers have discussed the issue of privacy. Bae and Kim (2010) suggested that, in order to achieve a high level of privacy, the user should be given the authority to control the privacy settings when he/she receives or requests a service related to his/her personal information. The authors noted the importance of designing a privacy policy to protect personal information by blocking some people from seeing all or part of the user's personal information. They also designed a privacy model using mobile agents.

Dötzer (2006, p.4) stated that "once privacy is lost, it is very hard to re-establish that state of personal rights." This shows that privacy is essential to the construction of all communication systems, particularly mobile systems. The concept of self-representation enables users to interact and introduce themselves based on the data placed on profile pages such as name and pictures with others. Privacy is an important aspect of self-representation on online social networks since people share certain information with the public and receive information or comments from others.

The nature and complexity of the Internet cause some threats to web privacy (Bouguettaya&Eltoweissy 2003). According to Wang and Cui (2008), privacy is a state or condition of limited access to a person. Privacy regulations can be defined as a set of rules or policies set by users to achieve a certain level of privacy. In terms of location privacy, privacy regulations restrict access to information on a user's location. Each privacy rule or policy can include some restrictions (Sadeh& Hong 2009).

Although there is no policy mandating online personal information privacy, some types of privacy solutions do exist (Passant et al. 2009). These solutions can be classified into: protective technologies, social awareness and legislative support. Protective technologies, such as strong authentication and access control, have developed quickly and have evolved over time. These rely on encryption as a way to solve privacy concerns. The second type of solution, social awareness, involves educating people about the possible risks of personal information misuse when they provide data such as their home address and mobile phone number. Lastly, legislation can be enacted to clarify aspects of the agreement with users to protect the collection of personal information under the framework of the law (Campisi, Maiorana&Neri 2009).

## 2.3. Online social network privacy

Social network users can share different pieces of personal information with others; however, these details may be misused by friends. Gross and Acquisti's study (2005) of Facebook users' privacy concerns found that 91 percent of users uploaded their pictures, 88 percent shared their date of birth, 40 percent showed their phone number, and 51 percent wrote their current address. Sharing personal information such as this can lead to misuse of data, whether intentional or not. For example, some people share profile details such as their full name, gender and phone number with their friends. If the social network account of one of the user's friends is hacked, the spammer or the hacker can misuse these details to blackmail the user (Rosenblum 2007). Another example is the misuse of data on relationship status. If user X is engaged to user Y, and user X hides his/her relationship status from his/her profile but user Y does not, then other users who are able to see his/her profile details can see the relationship status for user X through user Y's profile (Gundecha, Barbier& Liu 2011).

A study by Williams et al. (2009) found that older users are more careful about posting personal information details such as date of birth, friend list and school information on their social network

International Journal of Security, Privacy and Trust Management ( IJSPTM) vol 2, No 2, April 2013

accounts than younger users. Similarly, Zukowski and Brown (2007) found that older Internet users are more concerned about the privacy of personal information than younger users. Some of the younger users publish their own information without the knowledge of the risks that may occur from the misuse of these data. The online environment may be dangerous for users, regardless of age, because of the possible leakage of personal information details.

George (2006) cited the case of US college athletes whose pictures, which they posted online, were misused by a website, which publishes stories about scandals in sport. The author pointed out that the issue of privacy has not gone unnoticed by social network providers.
Gross and Acquisti (2005) conducted a study on a sample of 4,000 students from Carnegie Mellon University who use social network accounts. They found that a large proportion of students did not care about the privacy risks that might increase the chance of a third party misusing a student's personal information. Another study by the same authors claimed that more than 77 percent of the respondents did not read privacy policies (Acquisti& Gross 2006).

The ability to control privacy options is essential to increasing the users' confidence in their social network providers. Since Internet users represent a range of different cultures and ages, privacy options should be clear, simple and easy to use. Users must have the ability to control their privacy options at any time. These privacy options allow users to accept or reject the dissemination of their information to others. For example, some users do not want to publish sensitive information such as health or medical information (Samavi&Consens 2010). These users are aware that people with less than honourable intentions can harm adults or children by misusing their personal information. A study conducted by Casarosa (2010) found that minors are interested in new technologies and the Internet, and can be contacted by strangers online asking to form a friendship. When a website publishes the personal information of a minor without giving the child's parents (or the child's guardian) the authority to select privacy options, potential predators can use some of the minor's personal information, such as a mobile phone number, to engage in sexual contact (Casarosa 2010).

## 2.4. Online privacy risks

There are several risks surrounding the posting of personal information details on social networks. These threats can be caused by hackers or spammers who obtain users' personal information details. Identity theft is one of the major risks that users face (Williams et al. 2009). Access to sensitive information may also lead to terrorism risks, financial risks and physical or sexual extortion (Gharibi&Shaabi 2012).

Gao et al. (2011) discussed the common privacy breach attacks in online social networks. First, users usually upload their personal information when they trust the service provider. However, the provider can use these details for business purposes such as advertising. In addition, it is not only the service providers who can see the users' personal information. Some online social networks provide users with policies to determine the list of authorized persons who can see their personal information. These policies vary from one provider to another; some providers give users more flexibility than others and some provide encryption for their data. The second privacy breach can be caused by the user's friends, who can share the user's personal information details with others. Friends who have access to the user's personal information can copy and publish this information. The third breach is due to spammers. When spammers see the user's friend list, they can see other users' personal information by sending them a friend request, impersonating one of his/her friends by using the friend's name or picture. Lastly, breaches can be caused by third party applications installed by users. These applications can be a threat to users, especially if they are not from a trusted provider. When the application accesses the users' personal information, others can obtain this information.

Novak and Li (2012) also claimed that privacy breaches can be caused by friends, applications, and exploitation of personal information details by the service providers for advertising. The authors added that understanding privacy settings is not enough to protect users, especially from friends and other online social network users. Thus, social networking websites such as Facebook prioritize the development of tools to protect privacy. This is manifested in the social network providers' requests for new users to create new privacy settings. However, some users do not realize the risk of leakage of personal information (Lee et al. 2011). Therefore, sensitive information such as home address and date of birth should not be published online in order to avoid risks to online privacy. Increasing user awareness of these risks, providing a privacy management system for users to control their personal information details, and constantly updating privacy policies can lead to a decline of these risks (Gharibi&Shaabi 2012).

## 2.5. Online privacy protection

Privacy settings, that allow the user to control the profile view and distribution of personal data, vary across social networking websites, and there is no privacy standard for controlling the user's personal information settings. Although privacy settings should be chosen carefully, most online social network providers have complex privacy settings (Novak & Li 2012). These complex privacy settings may cause confusion among users (Gundecha, Barbier& Liu 2011).

Different techniques have been designed to increase personal information privacy protection. Williams et al. (2009) listed some steps for online social network users to stay safe. These included: being aware of the risks of social networks, limiting the posting of personal information details, and being careful when dealing with strangers online or when reading any information from any sender.

Most social networking sites have given their users more authority to control privacy settings. Users of some social networking sites are now able to classify their friend list into sub-lists, which allows some personal information details, such as birthday or relationship status, to be visible to one sub-list and hidden from others. Fang et al. (2010) designed a privacy recommendation wizard based on user inputs to help users classify their friend list into sub-lists. The wizard gives the user two options: to allow the friends in their sub-list to see their personal information, or to deny them access to this information.

Configuring privacy settings so that only friends can see your posts is not enough to defend oneself from other attacks such as applications and advertisements (Stutzman& Kramer-Duffield 2010). Lipford, Besmer and Watson (2008) found that showing an example of privacy settings will enable users to understand their privacy settings better. It will also help them find out who can see their personal information. For instance, Facebook allows users to see their profiles from their friends' point of view; this allows users to see what personal information their friends can see. This technique helps users understand privacy settings but does not provide security to protect them from neighbourhood attacks or other types of attacks such as viruses or spam.

Fang and Lefevre (2010) designed a privacy wizard system to make it easier for users to control their privacy settings. It was designed based on friends classification to groups and asking questions. It guides users in choosing privacy settings for group or individual users by allowing the users to see or hide an item. For example, if Alice is Bob's friend, then Bob can identify which items of his profile Alice can see. Bob can hide some details such as his date of birth and mobile number from Alice, and can do the same for his other friends.

## 2.6. Trust

Trust is an important element that can determine the success of online social networks and other business websites. Social network providers and other service providers have to increase the users' confidence and push them to trust their services by providing a secure and easy system for users' personal information privacy protection. Spam removal applications are the most trusted applications for users (Hameed et al. 2011; Grier et al. 2010). Users may also trust communications between their computers and the internet websites more than online social network providers in terms of leakage of personal information (Cutillo, Molave&Strufe 2009).

PayPal is an example of a trusted system. It is used to complete electronic transactions between the seller and the buyer, both of whom have to trust the system. Each user needs to create an account using authentic personal information and credit card details (Lutz 2012). One of the main reasons for the customers' confidence in the electronic payment system is their trust in the service provider's ability to maintain privacy and security (Ally, Teleman& Cater-Steel 2010).

However, this study differs from most prior studies on personal information privacy in three key respects. First, measure the use of mobile web services such as email, chat and social networks through smart mobile phones. Second, measure users' awareness of their social network accounts privacy settings and define the important personal information items. Three, this study examined both usability and suitability of smart mobile phones to control privacy settings.

## 3. OBJECTIVES

This study aims to measure the importance of privacy for Internet and mobile phone users and to define which pieces of personal information they consider more important in terms of privacy protection. It also aims to determine the awareness of users on privacy settings offered by social network providers, as well as their level of satisfaction with these settings. Lastly, the study aims to identify difficulties in choosing privacy settings on mobile phones, and to determine whether a more suitable method of choosing privacy settings on mobile phones needs to be developed. This information can be used to develop a smart wizard system which will help users choose suitable privacy settings easily and quickly from any device, including mobile phones.

International Journal of Security, Privacy and Trust Management ( IJSPTM) vol 2, No 2, April 2013

## 4. METHODOLOGY

### 4.1. Respondents of the study

The study sample was selected from the University of Dammam in Saudi Arabia and the New England University in Australia. The criteria for selecting respondents were that respondents should have at least one social networks account and mobile phone. In addition, the study population was both students and staff of University of New England and University of Dammam and aged 18 and more than 45 years. Data were collected from some departments and colleges through hard copy forms included with the participation invitations. Most of the participants from the University of Dammam were males because there is no mix between genders on colleges.

### 4.2. Survey

The survey questionnaire was available in two languages: English and Arabic. A total of 185 respondents completed the survey (95 used the Arabic questionnaire and 90 used the English questionnaire) and all questions were multi-choices questions.

## 5. SURVEY RESULTS

### 5.1. Summary of descriptive statistics

A total of 185 participants (157 males and 28 females) participated in the survey. Majority (75%) of the respondents belonged to the 18-25 age group. Table 1 shows that nearly all of the respondents had their own mobile phones, and most of them (92%) used their mobile phones to browse the Internet. Half (50%) of them used their mobile phones to browse the Internet at home. About 70 percent used both their mobile phones and computers to browse the Internet.

| Question | No. of respondents | Percent % |
|---|---|---|
| **Do you have a mobile phone?** | | |
| Yes | 184 | 99.5 |
| No | 1 | 0.5 |
| **Do you use your mobile phone for browsing the Internet?** | | |
| Always | 70 | 37.8 |
| Sometimes | 100 | 54.1 |
| Never | 15 | 8.1 |
| **Normally, where do you use mobile phone to browse the Internet?** | | |
| In the car | 12 | 6.5 |
| At home | 92 | 49.7 |
| At work | 11 | 5.9 |
| Other | 37 | 20.0 |
| More than 1 place | 32 | 17.3 |
| **What do you use to browse the Internet?** | | |
| Mobile phone | 5 | 2.7 |
| Computer | 51 | 27.6 |
| Both mobile phone and computer | 128 | 69.2 |

Table 1: Ownership of mobile phones and devices used to browse the Internet

### 5.2. Online social network accounts

To understand the behaviour of online social network users, participants were asked about their social network account usage, the amount of time spent browsing their account, and the number of friends on their friend list. Figure 1a shows that more than 45 percent of the respondents had more than one social network account, and 32 percent had Facebook accounts. Also, about 10 percent of the respondents had about 10 online social networks friends whereas about 50 percent of them have more than than 100 friends (figure 1b). Nearly half (48%) of the users visited their accounts one to five times per day (Figure 1c). Forty percent of the respondents spent about 30 minutes per day browsing their accounts; and about 50 percent of them spent more than 30 minutes doing so (Figure 1d).

Figure 1e shows the rapid increase in the number of users with social network accounts. Thirty-seven percent of the respondents have had social network accounts for more than three years; in the past three years alone, about 61% created social network accounts. This increase may be due to many factors, one of which is the widespread use of mobile phones to browse the Internet.

## 5.3. Mobile web services

Results show that participants use their mobile phones for a variety of uses. Most of the respondents use mobile web services such as email, chat and social networks. On a scale of 1 to 5 (1 means never, and 5 means always) the mean scores ranged from 2.75 (for checking news and weather) to 3.23 (for accessing email) (See Figure 2). This confirms that mobile phones are not only used to make phone calls, but also to access other services. Because of the annual increase in the number of social network users and the number of people who use mobile phones to browse the Internet, mobile web services should be improved to be easier for browsing and controlling settings, especially for small mobile phones screens.

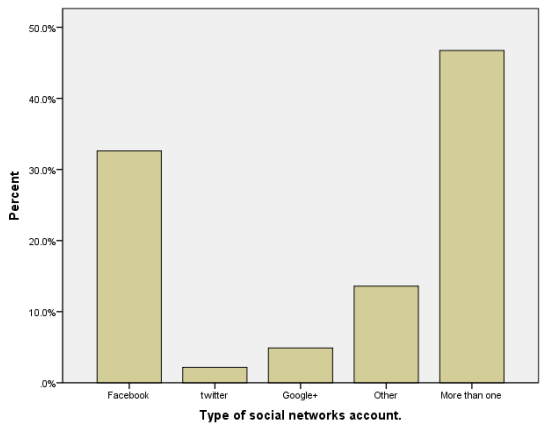
Figure1a: Distribution of online social networks accounts.

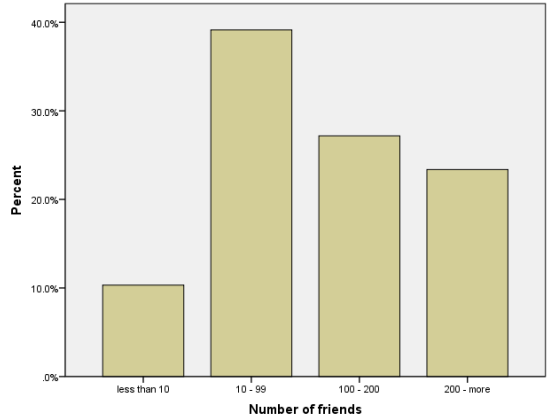
Figure 1b: Distribution of the members' number of friend list.

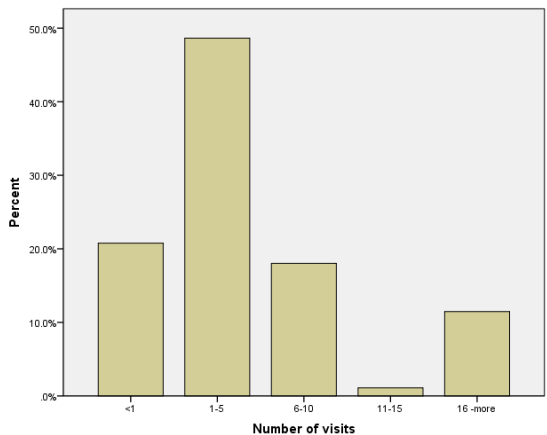
Figure 1c: Average number of visits per day.

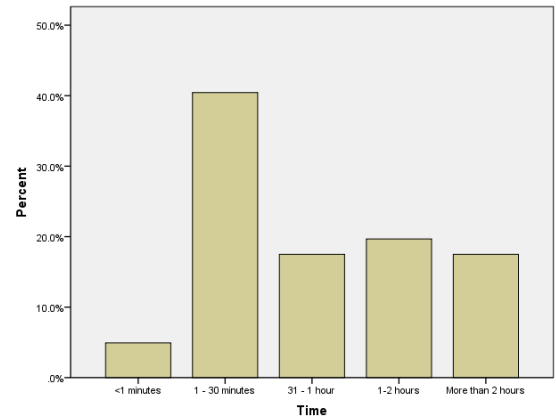
Figure 1d: Average spending time for browsing the account.

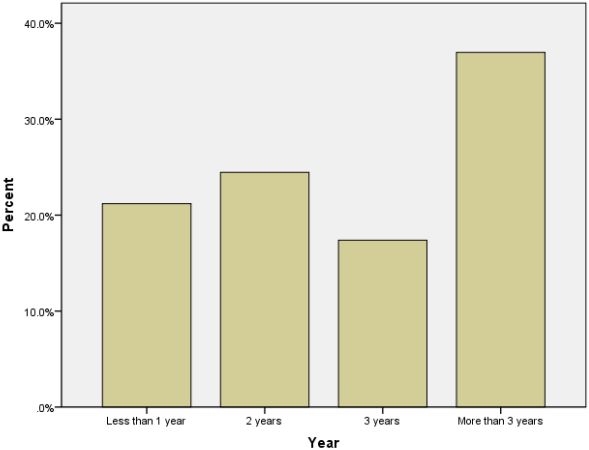

Figure 1e: Amount of time social network account has been activated.

Figures1a to 1e: Social network account usage

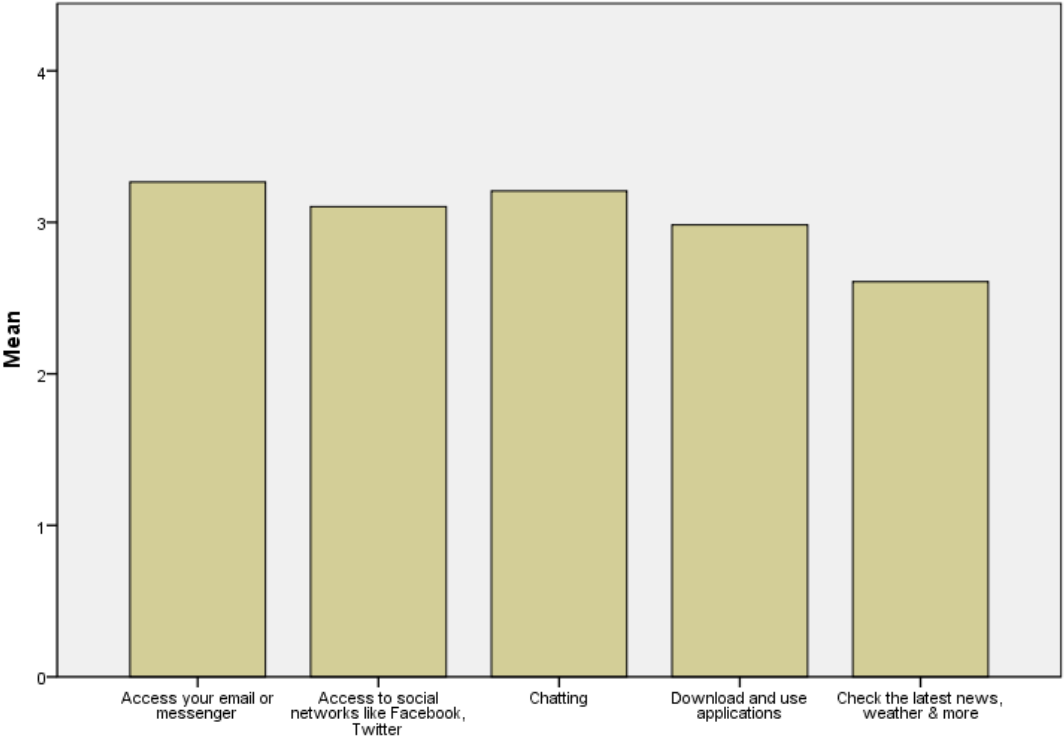

Figure 2: Use of mobile web services

## 5.4. Privacy settings

The survey questionnaire also contained questions on users' awareness of their social network account's privacy settings. Results of the survey show that although most of the users are aware of the privacy settings, most of them do not change their privacy settings from default settings (Table 2). Majority (67%) of the respondents were interested in controlling their privacy settings, but only 60 percent changed them. Only 40 percent changed their privacy settings regularly. Two-thirds (66%) of the users said they were familiar with privacy settings, and about three-fourths (73%) said they could prevent others from seeing their personal information. Majority (71%) of the respondents claimed they were completely satisfied with the method of selecting their account's privacy settings.

| Question | Yes % | No % | I don't know % |
|---|---|---|---|
| Are you interested in controlling the privacy settings of your account? | 67.0 | 23.8 | 9.2 |
| Have you changed the privacy settings of your account? | 59.5 | 35.1 | 4.9 |
| Are you familiar with your privacy settings? | 65.9 | 25.9 | 3.2 |
| Do you regularly change your privacy settings? | 40.0 | 56.8 | 3.2 |
| Are you completely satisfied with the method of selecting the privacy settings of your account? | 71.4 | 18.4 | 10.2 |
| Can you prevent other users from seeing your personal information? | 73.0 | 19.5 | 7.5 |

Table 2: Awareness and use of privacy settings

## 5.5. Risks of leaking personal information

The survey questionnaire also measured the respondents' awareness of the risks that could occur from the leakage of personal information. Respondents were asked if they posted real personal information on their accounts. Table 3 shows that two-thirds (66%) of the respondents were worried about the misuse of their personal information on their accounts. In addition, 69% of the respondents did not want strangers to see their personal information.

The respondents were also asked if their account providers shared their profile information with other websites. Slightly more than one-third (35%) said yes. This highlights the need to develop a framework that gives users the authority to allow or disallow websites to use their personal information.

Results show that users were cautious about accepting friend requests from strangers. Although 71 percent of the respondents received invitations to add an unknown person as a friend, 68 percent rejected these requests.

| Question | Yes % | No % | I don't Know % |
|---|---|---|---|
| Are you worried about the misuse of your personal information? | 66.5 | 26.5 | 7.0 |
| Does your account provider share your profile information with other websites? | 35.1 | 44.3 | 20.5 |
| Do you sometimes receive an invitation to add an unknown person as a friend? | 71.4 | 23.7 | 4.9 |
| Do you sometimes accept an invitation to add an unknown person as a friend? | 26.5 | 68.1 | 5.4 |
| Do you use real personal information in your account? | 64.3 | 33.5 | 2.2 |
| Do you want strangers to see your profile? | 23.8 | 68.6 | 7.6 |

Table 3: Awareness of the risk of personal information misuse

## 5.6. Using mobile phones to adjust privacy settings

The survey also measured the percentage of respondents who use a mobile phone to change the privacy settings of their social network accounts. While it is difficult to make comparisons between the more recent Internet mobile devices and older computers, there are some noticeable issues such as limitation of the screen in displaying the page content and flexibility of movement between the options. Some websites offered mobile web versions to facilitate browsing, but based on the results shown in Table 4, developing the current privacy systems for selecting personal information settings are required. Table 4 shows that only one-third (34%) of the respondents used their mobile phones to change their privacy settings. Slightly less than half (41%) of the respondents said that the size of their mobile phone screen was enough to allow control of the privacy settings. Only 42% of the respondents said that it was easy to change their account's privacy settings through their mobile phones.

| Question | Yes % | No % | I don't know % |
|---|---|---|---|
| Do you use your mobile phone to change the privacy settings? | 34.0 | 53.0 | 13.0 |
| Is the size of your mobile screen enough to allow control of the privacy settings? | 41.0 | 42.2 | 16.8 |
| Is it easy to change the privacy settings for your account through your mobile phone? | 42.2 | 36.8 | 21.1 |

Table 4: Use of mobile phones to control privacy settings

## 5.7. Importance of personal information details

As shown earlier, majority of the respondents are concerned about the privacy of personal information on their social network accounts. To determine which pieces of personal information they are most concerned about, respondents were asked to rank each item of personal information according to the importance of protecting the privacy of each item. The average rank (1 means low, and 5 means high) of each item of personal information is presented in Table 5.

Users consider contact information, such as current address, physical address, phone number and email address, as needing more privacy protection than other pieces of personal information. Items such as favourite TV shows, books and movies are not as important in terms of the need for privacy protection.

| Descriptive Statistics | | | | | |
| --- | --- | --- | --- | --- | --- |
| Item | N | Minimum | Maximum | Mean | Std. Deviation |
| Current address | 182 | 1 | 5 | 3.29 | 1.586 |
| Phone number | 182 | 1 | 5 | 3.21 | 1.656 |
| Physical address | 182 | 1 | 5 | 3.21 | 1.633 |
| Email | 184 | 1 | 5 | 3.15 | 1.432 |
| Friend list | 182 | 1 | 5 | 3.09 | 1.485 |
| Pictures | 182 | 1 | 5 | 3.07 | 1.556 |
| Videos | 182 | 1 | 5 | 3.02 | 1.573 |
| Education and work | 182 | 1 | 5 | 2.86 | 1.523 |
| Date of birth | 184 | 1 | 5 | 2.83 | 1.494 |
| Relationship status | 182 | 1 | 5 | 2.8 | 1.54 |
| Website | 181 | 1 | 5 | 2.78 | 1.533 |
| School information | 182 | 1 | 5 | 2.76 | 1.528 |
| Tags | 182 | 1 | 5 | 2.71 | 1.389 |
| Name | 184 | 1 | 5 | 2.63 | 1.597 |
| Comments and posts | 182 | 1 | 5 | 2.62 | 1.447 |
| Hometown | 182 | 1 | 5 | 2.52 | 1.448 |
| Religion | 182 | 1 | 5 | 2.52 | 1.72 |
| Gender | 184 | 1 | 5 | 2.48 | 1.533 |
| Favourite music | 182 | 1 | 5 | 2.42 | 1.513 |
| Interests and activity | 182 | 1 | 5 | 2.35 | 1.44 |
| Favourite movies | 182 | 1 | 5 | 2.35 | 1.478 |
| Favourite books | 182 | 1 | 5 | 2.26 | 1.448 |
| Favourite TV shows | 182 | 1 | 5 | 2.24 | 1.42 |

Table 5: Personal information details ranked according to importance of privacy protection

## 6. LIMITATION OF RESEARCH

The study suffers from the quality of the composition of the sample. Due to the majority of survey participants are males, the sample was not sufficiently heterogeneous. Also, by comparing the number of survey participants (both students and staff) with the number of both universities members, this might not represent the majority of the students and staff.
In addition, diversity of different cultures around the world is good idea for our study, but in this survey the cultural differences between both universities may affect the results, but this can help us to develop the smart wizard system supporting different cultures.

## 7. FUTURE WORK

As shown from survey results, high percentage of the use of smart phones for web services but the current privacy settings for online social networks need to be improved to support different type of mobile phones screens. So, there is a plan to develop a smart wizard system to help users in selecting personal information privacy settings and it will be designed based on this study.

## 8. CONCLUSION

This study focused on the importance of protecting social network users' personal information. It examined users' awareness of the risks and threats to their personal information privacy, and highlighted the need to develop a new privacy system supported by mobile Internet devices. Furthermore, survey results show high percentage of the use of smart phones for web services but the current privacy settings for online social networks need to be improved to support different type of mobile phones screens. Because most users use their mobilephones for Internet services, privacy settings that are compatible with mobile phones need to be developed. The method of selecting privacy settings should also be simplified to provide users with a clear picture of the data that will be shared with others.

Although there is no standard for controlling personal information privacy settings, it is recommended that social network providers design a system which protects users from different types of threats and risks. Privacy systems of online social networks, especially those which are used for selecting personal information settings, can be improved in several ways. Controlling privacy settings from different types of internet mobile devices and supporting various screens sizes for these devices, will help users to control their personal information privacy settings easily in different places and times. Furthermore, using a predictable wizard system for setting privacy settings is a suggestion to support different Internet mobile devices and different screen sizes of mobile phones. In addition, privacy systems should inform users about which items of personal information are displayed and which are not.

**Authors**

**NahierAldhafferi**
Nahier is a PhD student at University of New England in Australia. He works as a Lecturer at school of computer science in University ofDammam in Saudi Arabia. He holds a master degree in Internet Technology from University of Wollongong, Australia. Also, he is a lecturer at CISCO academy.

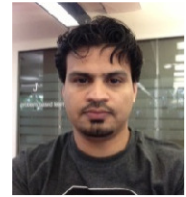

**Dr. Charles Watson**
Dr. Charles Watson is a UNE Council Member. He is aConvener of the Information Research Group and Faculty A&S Representative on e-ResearchWorking Group. Also he is a Computer Science Representative on School S&T Research Committee. He lectures Computer Science at the University of New England in Australia, including units in networks, security, e-commerce and forensic computing. He was previously a Senior Research Scientist at DSTO and CSIRO and holds a PhD in Computer Science from the University of Sydney.

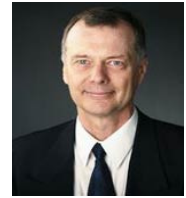

**Professor A. S. M. Sajeev**
Sajeev is the Chair in Computer Science/Information Technology at the University of New England. He holds a PhD in Computer Science from Monash University, Australia.His research interests are in software engineering: software metrics, testing, processes and project management. He also works in the area of mobile systems: interface design, language design, security and testing. He coordinates the IT Security theme within the University's targeted research area on security.

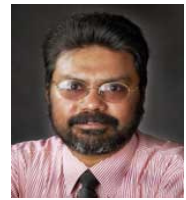